\documentclass[prd,twocolumn,showpacs,preprintnumbers,amsmath,amssymb,floatfix,nofootinbib]{revtex4}

\usepackage{subfigure,graphicx,epsfig,amsmath,amsfonts,amssymb,xcolor,slashed,ulem}

\usepackage{bm}
\usepackage{graphicx}
\usepackage{amsmath}
\usepackage{amsfonts}
\usepackage{amssymb}
\usepackage{color}
\usepackage{ulem}

\usepackage[section]{placeins}

\usepackage{booktabs}%¸Ä±í¸ñÏßÌõ´Öϸhline, ʹÓÃ\toprule, \midrule and \bottomrule
\usepackage[colorlinks, citecolor=blue,anchorcolor=red,menucolor=red, linkcolor=red,filecolor=red,runcolor=red,urlcolor=blue,frenchlinks=red]{hyperref}

\newcommand{\be}{\begin{equation}}
\newcommand{\ee}{\end{equation}}
\newcommand{\ba}{\begin{eqnarray}}
\newcommand{\ea}{\end{eqnarray}}
\newcommand{\nn}{\nonumber}

\newcommand{\mev}{\textrm{ MeV}}

\begin{document}
%\date{\today}
\bibliographystyle{unsrt}
\arraycolsep1.5pt

\title {Tau decay into $\nu_\tau$ and $a_1(1260)$, $b_1(1235)$, and two $K_1(1270)$}

\author{L.~R.~Dai}
\email[]{dailianrong68@l26.com}
\affiliation{School of Science, Huzhou University, Huzhou 313000, Zhejiang, China}
\affiliation{Department of Physics, Liaoning Normal University, Dalian 116029, China}
\affiliation{Departamento de F\'isica Te\'orica and IFIC, Centro Mixto Universidad de Valencia-CSIC,
Institutos de Investigac\'ion de Paterna, Aptdo. 22085, 46071 Valencia, Spain
}

\author{L. Roca}
\email[]{luisroca@um.es}
\affiliation{Departamento de F\'isica, Universidad de Murcia, E-30100 Murcia, Spain}

\author{E.~Oset}
\email[]{oset@ific.uv.es}
\affiliation{Departamento de F\'isica Te\'orica and IFIC, Centro Mixto Universidad de Valencia-CSIC,
Institutos de Investigac\'ion de Paterna, Aptdo. 22085, 46071 Valencia, Spain
}

\date{\today}
\begin{abstract}

We study the $\tau \to \nu_\tau A$ decay, with $A$ an axial-vector meson. We produce the $a_1(1260)$ and $b_1(1235)$ resonances in the Cabibbo favored mode and two $K_1(1270)$ states in the Cabibbo suppressed mode.
We take advantage of previous chiral unitary approach results where these resonances appear dynamically from the vector and pseudoscalar meson interaction in s-wave. Actually two different poles were obtained associated to the $K_1(1270)$ quantum numbers.
 We find that the unmeasured rates for  $b_1(1235)$ production are similar to those of the $a_1(1260)$ and for the two $K_1$ states we suggest to separate the present information on the $\bar K \pi \pi$ invariant masses  into $\bar K^* \pi$ and $\rho K$ modes, the channels to which these two resonances couple most strongly, predicting that these modes peak at different energies and have different widths. These measurements should shed light on the existence of these two $K_1$ states.

\end{abstract}

\maketitle

\section{Introduction}

Tau decays, with about 65 \% branching fraction into hadronic channels \cite{pdg}, (being the $\tau$ the only lepton with enough mass to decay into hadrons), have proved to be a good tool to learn about strong interactions at low energies \cite{Schael:2005am,Davier:2005xq,Braaten:1991qm,Lafferty:2015hja}. In particular, some of the decay modes have one resonance in the final state and we are concerned about the production of particular resonances which stand for a molecular interpretation. Concretely, in this work we are concerned about the production of axial vector resonances, $A$, this is, $\tau \to \nu_\tau A$. Given that in the $\tau \to \nu_\tau q \bar q$ at the quark level, the quarks are $d \bar u$ for the Cabibbo favored process, this defines the hadronic state with isospin $I=1$. Hence, we can obtain the $a_1(1260)$ $(1^-(1^{++}))$ and  the $b_1(1235)$ $(1^+(1^{+-}))$. For Cabibbo suppressed processes the initial quark state is $s \bar u$ and hence we produce a state with $I=1/2$ and strangeness. This is the $K_1(1270)$ $(1/2(1/2^+))$. An issue we wish to raise in this work is the fact that the chiral theories for the axial-vector mesons predict two states for $K_1(1270)$ \cite{Roca:2005nm,Geng:2006yb} and we evaluate the rates for decay into either state and suggest the  way to differentiate the two states in experiment.
   In chiral unitary theory, the axial vector mesons are generated from the interaction of vector mesons with pseudoscalars \cite{Lutz:2003fm,Roca:2005nm,Zhou:2014ila}. The production of an axial-vector meson in the tau decay  proceeds then in the following way: all possible pairs of vector-pseudoscalar are produced and then they are allowed to interact among themselves, and in this process the resonances are generated,  decaying later on in vector-pseudoscalar pairs or other channels.  From the microscopical point of view this is done from the original $q \bar q$ pair creation by means of hadronization, where an extra $\bar q q$ pair is created with the quantum numbers of the vacuum. A technical way to implement this step is done in Ref.~\cite{Dai:2018thd} using the $^3P_0$ model \cite{Micu:1968mk,LeYaouanc:1972vsx,close3P0} and we shall use some results from this work here.

    In the literature there are many works dealing with the production of vector-pseudoscalar in tau decays using different approaches. In Refs.~\cite{Li:1995aw,Li:1995tv} vector meson dominance is used while in
    Refs.~\cite{Volkov:2012gv,Volkov:2017arr,Volkov:2018zlr,Volkov:2017cmv,volkovparti,Volkov:2019cja,Volkov:2019jug,Volkov:2020tle}
the Nambu Jona Lasinio model (NJL)
\cite{Nambu:1961tp} is used. In those works the axial resonances when suited, are introduced explicitly via amplitudes dictated by symmetries \cite{Li:1995aw,Li:1995tv} or with explicit coupling to quarks in the NJL model.  This is different to our approach, since what we do is produce the vector-baryon pairs and then, using chiral dynamics and coupled channels Bethe Salpeter equations the pairs are allowed to interact and the interaction generates the axial-vector resonances, which are implicit in the scattering amplitudes used.

 The $\tau \to \nu_\tau a_1(1260)$ has been studied as part of the $\tau\to \nu_\tau \pi^+ \pi^- \pi^-$, which has had a wide attention experimentally \cite{Barate:1998uf,Nugent:2009zz,Briere:2003fr,Aubert:2007mh,Lee:2010tc,Nugent:2013hxa}. While the $a_1(1260)$ production provides the main contribution to the process, other mechanisms are at work when one wishes to get a very good agreement with experiment \cite{Nugent:2013hxa,Ivanov:1989qw,Bowler:1987bj,Isgur:1988vm,Colangelo:1996hs,Dumm:2009va,Was:2015laa,Sanz-Cillero:2017fvr,Guo:2008sh,
 Wagner:2008gz,Wagner:2007wy,Osipov:2018lnl}. 
  Special emphasis in the role of three body unitarity of the final three pion state is made in a recent paper \cite{Mikhasenko:2018bzm}.
  The $a_1(1260)$ dominance is seen in the $\pi \rho$ decay channel with the $\rho$ identified from the $\pi \pi$ mass distribution.

  Contrary to the $a_1(1260)$ production, which has been widely studied, the $b_1(1235)$ production, to the best of our knowledge has not been discussed, neither experimentally nor theoretically. An interesting aspect of the approach we follow is that, since the vector-pseudoscalar channels produced and their weights are very well defined in the tau decay, we can evaluate the production of the different axial-vector mesons  with the same approach. This is our aim here, and we shall relate the production of the $a_1(1260)$, $b_1(1235)$ and the two $K_1(1270)$ states. We shall see that the $b_1(1235)$ production rate is of the same order as the $a_1(1260)$ and the two $K_1(1270)$ states are produced with smaller rates, since they are Cabibbo suppressed by a factor about $\tan \theta_c^2\simeq 1/20$, but they have very distinct decay modes, which we propose to differentiate.
    The $K_1(1270)$ production has also been reported in the PDG \cite{pdg,Barate:1999hj}, but the $K_1(1270)$ is identified by looking at the $K \pi \pi$ mass distribution, which contains both $K^*\pi$ and $\rho K$. According to \cite{Geng:2006yb} the two $K_1(1270)$ resonances couple very differently to these two modes, and we suggest that the two modes are separated to visualize the two resonances peaking at different energies and with different widths.   Theoretically, the production of the $K_1(1270)$ has also been addressed in \cite{volkovparti,Volkov:2018zlr} from the perspective of the Nambu Jona Lasinio model, but only one $K_1$ state is considered there.

    In the present work we shall use results of Ref.~\cite{Dai:2018thd} and relate the production rates of the four axial vector resonances. The approach of \cite{Dai:2018thd} does not calculate absolute rates but just ratios between different production channels. Here we follow the same strategy and relate all the rates with the production of the $a_1(1260)$ resonance. In addition we calculate partial decay rates into different channels to facilitate the experimental work identifying the production of the $b_1(1235)$ and the two $K_1(1270)$ states.

\section{Formalism}

\subsection{Microscopical formalism for meson meson production}

\begin{figure}[h]
\begin{center}
\includegraphics[width=0.5\textwidth]{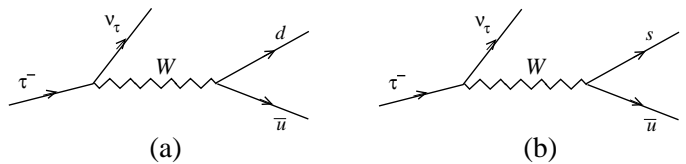}
\caption{\small{(a) Cabibbo favored $\tau^-$ decay to quark-antiquark. (b) Cabibbo suppressed decay}}
\label{fig:Tquarks}
\end{center}
\end{figure}

Let us begin with the elementary process at the quark level for the $\tau$ decay, $\tau\to\nu_\tau d\bar u$ for the Cabibbo favored mode and $\tau\to\nu_\tau s\bar u$ for the Cabibbo suppressed mode.
 They are depicted in
 Fig.~\ref{fig:Tquarks}

In order to create a vector and a pseudoscalar meson we introduce a $\bar q q$ pair with the quantum numbers of the vacuum. This was done in Ref.~\cite{Dai:2018thd} using the
$^3P_0$ model. Before one looks into the dynamics of the weak decay, it is easy to see which pairs are created and with which weight. Following Ref.~\cite{Dai:2018thd} we write the matrix $M$ for $q\bar q$,

\begin{equation}
M\equiv \left(\begin{array}{ccc}u\bar u & u\bar d & u\bar s\\
					       d\bar u & d\bar d & d\bar s\\
					       s\bar u & s\bar d & s\bar s
\end{array}\right)\,.
\label{eq:Mqqbar}
\end{equation}

\begin{figure}[h]
\begin{center}
\includegraphics[width=0.4\textwidth]{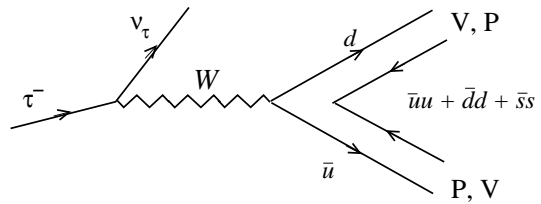}
\caption{\small{Hadronization of the primary $q \bar q$ pair to produce a vector and pseudoscalar meson.}}
\label{fig:TquarksVP}
\end{center}
\end{figure}
The hadronization of the $d\bar u$ pair is depicted  in Fig.~\ref{fig:TquarksVP} and proceeds as

\begin{align}
d\bar u \to
\sum_{i=1}^3 d \bar q_i q_i \bar u=
 \sum_{i=1}^3{M_{2i}M_{i1} } =(M^2)_{21}.
\label{eq:qqhadr}
\end{align}

Next we can identify the hadronic states produced with the physical pseudoscalar and vector mesons associating the  $M$ matrix to the $SU(3)$ matrices containing the pseudoscalar and vector mesons:

\begin{align*}
M\Rightarrow P\equiv
\left(\begin{array}{ccc}
              \frac{\pi^0}{\sqrt{2}}  + \frac{\eta}{\sqrt{3}}+\frac{\eta'}{\sqrt{6}}& \pi^+ & K^+\\
              \pi^-& -\frac{1}{\sqrt{2}}\pi^0 + \frac{\eta}{\sqrt{3}}+ \frac{\eta'}{\sqrt{6}}& K^0\\
              K^-& \bar{K}^0 & -\frac{\eta}{\sqrt{3}}+ \frac{2\eta'}{\sqrt{6}}
      \end{array}
\right)\,,
\label{eq:Pmatrix}
\end{align*}
which makes use of the standard $\eta$ and $\eta'$ mixing \cite{Bramon:1992kr}, and

\begin{equation}
 M\Rightarrow V \equiv
 \left(\begin{array}{ccc} \frac{1}{\sqrt{2}} \rho^0 +
\frac{1}{\sqrt{2}}\omega
 & \rho^+ & K^{*+}\\
\rho^-& - \frac{1}{\sqrt{2}} \rho^0 + \frac{1}{\sqrt{2}}\omega
& K^{*0}\\
K^{*-}& \bar{K}^{*0} & \phi
\end{array}
\right). \label{eq:Vmatrix}
\end{equation}

Then we get the contributions

\begin{align}\label{eq:PV21}
(P \cdot V)_{21}=&\pi^- \left( \frac{\rho^0}{\sqrt{2}}+\frac{\omega}{\sqrt{2}}  \right) \nn \\
+&
\left(-\frac{\pi^0}{\sqrt{2}}+\frac{\eta}{\sqrt{3}}+ \frac{\eta^{\prime}}{\sqrt{6}} \right)\rho^{-} + K^0 K^{*-} \, , \nn \\
(V \cdot P)_{21}=&\rho^- \left(\frac{\pi^0}{\sqrt{2}}+\frac{\eta}{\sqrt{3}}+ \frac{\eta^{\prime}}{\sqrt{6}} \right) \nn \\
 + &\left( -\frac{\rho^0}{\sqrt{2}}+\frac{\omega}{\sqrt{2}}  \right)\pi^- + K^{*0}K^- \, .
\end{align}
It is important to keep track of the order because the weak transition has two terms and one of them changes sign when changing the order of $PV$ to $VP$. We shall see that this is directly tied to the $G$-parity of the states produced.
For the Cabibbo suppressed case we proceed analogously but the  hadronization of $s\bar u$ gives rise to the matrix element $(M^2)_{31}$ with the result

\begin{align}\label{eq:PV31}
(P \cdot V)_{31}=&K^- \left(\frac{\rho^0}{\sqrt{2}}+\frac{\omega}{\sqrt{2}}\right)+\bar{K}^0 \rho^-\nn \\
+&\left(-\frac{\eta}{\sqrt{3}}+\frac{2\eta^{\prime}}{\sqrt{6}} \right)K^{*-} \, ,\nn \\
(V \cdot P)_{31}=&K^{*-} \left(\frac{\pi^0}{\sqrt{2}}+\frac{\eta}{\sqrt{3}}+\frac{\eta^{\prime}}{\sqrt{6}}\right)+\bar{K}^{*0}\pi^- +\phi K^{-}  \, .
\end{align}

The next ingredient is the evaluation of the weak interaction operators. The weak interaction is given by the product of the lepton and quark currents
\begin{eqnarray}
H= \mathcal{C} L^\mu Q_\mu,
\label{eq:LQ}
\end{eqnarray}
where in $\mathcal{C}$, which we take constant, we include weak interaction couplings and radial matrix elements of the quark wave functions. These matrix elements are smoothly dependent on the momentum transfer given the small phase space for the reactions, and then, in ratios of rates in which we are interested, the factor $\mathcal{C}$ cancels.
(Actually, the difference in the weak coupling between the Cabibbo allowed and supressed modes will be taken into account later on through the appropriate Cabibbo mixing angle, $\theta_c$).
In Eq.~\eqref{eq:LQ},  $L^\mu Q_\mu$ is the leptonic current

\begin{eqnarray}
L^\mu=\langle \bar u_\nu |\gamma^\mu- \gamma^\mu\gamma_5| u_\tau\rangle,
\end{eqnarray}
and $Q_\mu$ the quark current
\begin{eqnarray}
Q^\mu=\langle \bar u_d|\gamma^\mu-\gamma^\mu\gamma_5|v_{\bar u}\rangle.
\end{eqnarray}
We evaluate the matrix elements in the frame where the resonance is at rest. There $\gamma_5 v= \gamma_5 u$ and at these low energies only $\gamma^\mu \to\gamma^0\to 1$ and
 $\gamma^\mu \gamma_5 \to \gamma^i \gamma_5 \to \sigma^i$ survive and we get the components

\begin{eqnarray} \label{eq:Qu2}
Q_0&=& \langle \chi^{\prime} | 1 | \chi \rangle \equiv M_0  \nonumber \, , \\
Q_i&=&  \langle \chi^{\prime} | \sigma_i |\chi  \rangle \equiv N_i \, ,
\end{eqnarray}
with $i=1,2,3$, where $\chi$, $\chi^{\prime}$, are Pauli spinors.

The other point to consider is that, as discussed in \cite{Dai:2018thd}, the $VP$ mesons are produced in s-wave, unlike $PP$ modes which would be produced
in p-wave. This forces the $d\bar u$ pair to be produced
in $L=1$ to have positive parity at the end. With the $G$-parity rule $G=(-1)^{L+S+I}$ ($L=1$, $I=1$, $S=0$ for the operator ''1" and 1 for the operator $\sigma_i$) we see that $M_0$ carries $G$-parity positive while $N_i$ carries negative $G$-parity.

The direct calculation of matrix elements done in \cite{Dai:2018thd} using Racah algebra to produce the final meson pair with total angular momentum states $|JM\rangle$ and  $|J'M'\rangle$, gives

\begin{align}
M_0(PV)&=M_0(VP)=\frac{1}{\sqrt{6}}\frac{1}{4\pi}\, ,  \nn \\
N_\mu(PV)&= -(-1)^{-\mu} \frac{1}{\sqrt{3}}\frac{1}{4\pi} {\cal C}(1 1 1; M',-\mu,M'-\mu)\,\delta_{M0}\, , \nn \\
N_\mu(VP)&= (-1)^{-\mu} \frac{1}{\sqrt{3}}\frac{1}{4\pi} {\cal C}(1 1 1; M,-\mu,M-\mu)\,\delta_{M'0}\, .
\label{eq:M0Nnu}
\end{align}
It is important to keep in mind tha $N_i$ changes sign from the combination $PV$ to $VP$. This fact is essential to recover the $G$-parity conservation
 in the final state interaction, as we shall see. With this rule in mind and
 Eqs.~\eqref{eq:PV21} and \eqref{eq:PV31} one easily finds the weights $h'_i$, $\bar h'_i$ that multiply the $M_0$ and $N_i$ operators for each of the coupled channels. These factors are shown in Tables~\ref{tab:ha1}-\ref{tab:hK1}.

%\begin{widetext}
\begin{center}
\begin{table}[h!]
\begin{center}
\caption{\small{Weights $h'_i$ ($\bar h'_i$) of the different $I=1$, $G=-1$, $VP$ components for  the $M_0$ ($N_i$) amplitudes.}}
\begin{tabular}{|c|c|c|c|c|}
\hline
   & $\rho^0\pi^-$ & $\rho^-\pi^0$   & $K^{*0} K^{-} $   &  $K^{*-} K^{0} $    \\ \hline $h'_i$      &  0  & 0 & 1 & 1      \\ \hline
$\bar h'_i$ &  $\sqrt{2}$    & $-\sqrt{2}$ & -1 & 1        \\
 \hline
\end{tabular}
\label{tab:ha1}
\end{center}
\end{table}
\end{center}
%\end{widetext}

%\begin{widetext}
\begin{center}
\begin{table}[h!]
\begin{center}
\caption{\small{
Same as Table~\ref{tab:ha1} but for $I=1$, $G=+1$.}}
\begin{tabular}{|c|c|c|c|c|}
\hline
     & $\omega\pi^-$ & $\rho^-\eta$   & $K^{*0} K^{-} $   &  $K^{*-} K^{0} $    \\ \hline
      $h'_i$      &  $\sqrt{2}$  & $\frac{2}{\sqrt{3}}$ & 1 & 1      \\ \hline
$\bar h'_i$ &  0    & 0 & -1 & 1        \\
 \hline
\end{tabular}
\label{tab:hb1}
\end{center}
\end{table}
\end{center}
%\end{widetext}

\begin{widetext}
\begin{center}
\begin{table}[h!]
\begin{center}
\caption{\small{Same as Table~\ref{tab:ha1} but for  $I=1/2$.($\theta_c$ is the Cabibbo angle, $\tan \theta c=0.2312$)}}
\begin{tabular}{|c|c|c|c|c|c|c|c|}
\hline
     & $K^{*-} \pi^{0} $ & $\bar K^{*0}\pi^{-} $   & $\rho^0 K^{-} $   &  $\rho^- \bar K^{0} $ & $\omega K^{-}$ &  $\phi K^{-}$ & $K^{*-} \eta $ \\ \hline
 $h'_i$  &  $\frac{1}{\sqrt{2}}\tan\theta_c$  & $\frac{1}{\sqrt{2}}\tan\theta_c$  & $\frac{1}{\sqrt{2}}\tan\theta_c$ & $\tan\theta_c$ & $\frac{1}{\sqrt{2}}\tan\theta_c$ & $\tan\theta_c$ & 0 \\ \hline
 $\bar h'_i$  &  $-\frac{1}{\sqrt{2}}\tan\theta_c$  & $-\frac{1}{\sqrt{2}}\tan\theta_c$  & $\frac{1}{\sqrt{2}}\tan\theta_c$ & $\tan\theta_c$ & $\frac{1}{\sqrt{2}}\tan\theta_c$ & $-\tan\theta_c$ &$-\frac{2}{\sqrt{3}}\tan\theta_c$ \\ \hline
\end{tabular}
\label{tab:hK1}
\end{center}
\end{table}
\end{center}
\end{widetext}

As shown in \cite{Dai:2018thd}, the differential width for $PV+VP$ production is given by

\begin{align}\label{eq:dGdM}
\frac{ d\Gamma}{dM_{\rm inv}(M_1 M_2)} =  \frac{2\,m_\tau 2\, m_\nu}{(2\pi)^3} \frac{1}{4 m^2_\tau}\, p_\nu {\widetilde p_1}\, \overline{\sum} \sum \left|t\right|^2 \,,
\end{align}
where $p_\nu$ is the neutrino momentum in the $\tau$ rest frame,
and ${\widetilde p_1}$ the momentum of $M_1$ in the $M_1 M_2$ center of mass frame, and

\begin{align}\label{eq:ff2}
\overline{\sum} \sum \left|t\right|^2&=  \frac{1}{m_\tau m_\nu} \left(\frac{1}{4\pi}\right)^2
 \left[\left(E_\tau E_\nu + {\bm{p}}^2 \right) \frac{1}{2}{|h'|}^2_i \right.\nn \\
 &  \left.+  \left(E_\tau E_\nu -\frac{ {\bm{p}}^2 }{3}\right) |\bar{h}'|^2_i \right]
\end{align}
where $p$ is the momentum of the neutrino in the rest frame of $M_1 M_2$ and $E_\tau$, $E_\nu$, the energies of the $\tau$ and $\nu_\tau$
 for this momentum.

\subsection{Final state interaction}

\begin{figure}[h]
\begin{center}
\includegraphics[width=0.45\textwidth]{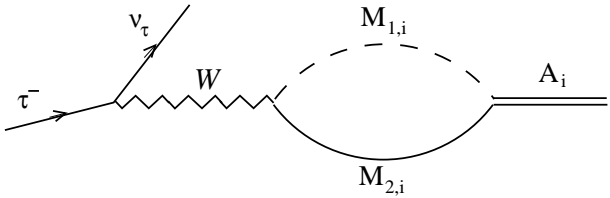}
\caption{\small{
Mechanism for the production of the dynamically generated axial-vector resonance $A_i$ through the coupled channels $M_{1,i}$, $M_{2,i}$, of pseudoscalar and vector mesons.
}}
\label{fig:TFSI}
\end{center}
\end{figure}

As mentioned in the Introduction, the axial-vector resonances are generated in our approach by the interaction of the mesons produced in a first step. Thus we have the mechanism shown in Fig.~\ref{fig:TFSI}

The dynamics for the $W M_1 M_2$ vertex has been discussed above and the new ingredients are the $M_1 M_2$ loop $G$ function and the couplings of the $A_j$ resonance to the $(M_1 M_2)_i$ channels. The information of the $G$ functions is given in \cite{Roca:2005nm,Geng:2006yb} and the couplings are given in Table~VII of \cite{Roca:2005nm} for the $a_1(1260)$ and the $b_1(1235)$ resonances, which are obtained from the residues  at the pole positions of the $VP$ scattering amplitudes in the proper unphysical Riemann sheets. The couplings
 for the two $K_1$ states are given in Table~IV in \cite{Geng:2006yb}.  The position of the poles can be associated with the  mass (real part) and width (twice the imaginary part) of the resonances and for the two $K_1(1270)$ states found, that we will call $K_1(A)$ and $K_1(B)$, they are:
 \begin{align}
 K_1(A)&:\quad M=1195\mev ,\quad \Gamma=246\mev, \nn\\
 K_1(B)&:\quad M=1284\mev ,\quad \Gamma=146\mev. \nn
\end{align}
 The first state, $K_1(A)$, couples dominantly to $ K^* \pi$ while the second one, $K_1(B)$, couples  mostly to $\rho K$. The couplings in the Tables of Refs.~\cite{Roca:2005nm,Geng:2006yb} are given for the $VP$ isospin states, which are related to the charge states by

 \begin{align}
&|\rho \pi; I=1, I_3=-1\rangle= \nn \\
& \qquad\qquad\qquad \frac{1}{\sqrt{2}} |\rho^0\pi^-\rangle- \frac{1}{\sqrt{2}}|\rho^-\pi^0\rangle \nn \\
& |\rho \bar K; I=1/2, I_3=-1/2\rangle= \nn \\
&  \qquad\qquad\qquad  -\frac{1}{\sqrt{2}} |\rho^0 K^-\rangle- \sqrt{\frac{2}{3}}|\rho^- \bar K^{0}\rangle \nn \\
 &   |\bar K^* \pi; I=1/2, I_3= -1/2\rangle= \nn \\
&  \qquad\qquad\qquad \frac{1}{\sqrt{3}} |K^{*-} \pi^{0}\rangle+ \sqrt{\frac{2}{3}}|\bar K^{*0}\pi^{-} \rangle\nn \\
&  |\bar K^* K + cc; I=1, I_3=-1,  G=+1  \rangle= \nn \\
&  \qquad\qquad\qquad  -\frac{1}{\sqrt{2}} \left(|K^{*-}K^{0}\rangle +|K^{*0} K^{-} \rangle  \right) \nn  \\
&   |\bar K^* K - cc; I=1, I_3=-1,  G=-1  \rangle=\nn \\
&   \qquad\qquad\qquad  -\frac{1}{\sqrt{2}} \left(|K^{*-} K^{0}\rangle -|K^{*0} K^{-} \rangle  \right)
    \end{align}
where   $\bar K^* K \pm cc$ actually stands for the $\pm 1 $ $G$-parity combination
$1/\sqrt{2}(K^* \bar K \pm \bar K^* K)$ \cite{Roca:2005nm}.
With this information we show in Tables~\ref{tab:ga1}-\ref{tab:gK1} the couplings of a given axial-vector resonance to the different $VP$ channels in terms of the couplings in the isospin basis of Refs.~\cite{Roca:2005nm,Geng:2006yb}.

%\begin{widetext}
\begin{center}
\begin{table}[h!]
\begin{center}
\caption{\small{Couplings of the  $I=1$, $G=-1$, axial-vector resonance $a_1(1260)$ to the different $VP$ channels}}
\begin{tabular}{|c|c|c|c|c|}
\hline
   & $\rho^0\pi^-$ & $\rho^-\pi^0$   & $K^{*0} K^{-} $   &  $K^{*-} K^{0} $    \\ \hline $g_i$      & $\frac{1}{\sqrt{2}}g_{a_1,\rho\pi}$  & $-\frac{1}{\sqrt{2}}g_{a_1,\rho\pi}$ & $\frac{1}{\sqrt{2}}g_{a_1,K^*\bar K}$ & $-\frac{1}{\sqrt{2}}g_{a_1,K^*\bar K}$      \\ \hline
\end{tabular}
\label{tab:ga1}
\end{center}
\end{table}
\end{center}
%\end{widetext}

%\begin{widetext}
\begin{center}
\begin{table}[h!]
\begin{center}
\caption{\small{Couplings of the  $I=1$, $G=+1$, axial-vector resonance $b_1(1235)$ to the different $VP$ channels}}
\begin{tabular}{|c|c|c|c|c|}
\hline
  & $\omega\pi^-$ & $\rho^-\eta$   & $K^{*0} K^{-} $   &  $K^{*-} K^{0} $    \\ \hline
  $g_i$      & $g_{b_1,\omega\pi}$  & $g_{b_1,\rho\eta}$ & -$\frac{1}{\sqrt{2}}g_{b_1,K^*\bar K}$ & $-\frac{1}{\sqrt{2}}g_{b_1,K^*\bar K}$      \\ \hline
\end{tabular}
\label{tab:gb1}
\end{center}
\end{table}
\end{center}
%\end{widetext}

\begin{widetext}
\begin{center}
\begin{table}[h!]
\begin{center}
\caption{\small{Couplings of the  $I=1/2$ axial-vector resonances $K_1(A)$ and $K_1(B)$ to the different $VP$ channels. The values of the couplings are different for both poles.}}
\begin{tabular}{|c|c|c|c|c|c|c|c|}
\hline
     & $K^{*-} \pi^{0} $ & $\bar K^{*0}\pi^{-} $   & $\rho^0 K^{-} $   &  $\rho^- \bar K^{0} $ & $\omega K^{-}$ &  $\phi K^{-}$ & $K^{*-} \eta $ \\ \hline
 $g_i$  &  $\frac{1}{\sqrt{3}} g_{K_1,\bar K^*\pi}  $  & $\sqrt{\frac{2}{3}} g_{K_1,\bar K^*\pi}$  & $-\frac{1}{\sqrt{3}}g_{K_1,\rho \bar K}$ & $-\sqrt{\frac{2}{3}} g_{K_1,\rho \bar K}$ & $g_{K_1,\omega \bar K}$ & $g_{K_1,\phi \bar K}$ & $g_{K_1,\bar K^*\eta}  $  \\ \hline
\end{tabular}
\label{tab:gK1}
\end{center}
\end{table}
\end{center}
\end{widetext}

The amplitude for the mechanism of Fig.~\ref{fig:TFSI} is readily obtained simply substituting

\begin{align}
h'_j\to h''_j=&\sum_i h'_i G_i(M_{A_j})g_{A_j,i} \nn \\
\bar h'_j\to \bar h''_j=&\sum_i \bar h'_i G_i(M_{A_j})g_{A_j,i}
\label{eq:hsubst}
\end{align}
where $i$ runs over the different coupled channels of each resonance $A_j$ and $g_{A_j,i}$ is the coupling of the $A_j$ resonance to the channel $i$. The third component of the vectors are easily taken into account. Indeed, the vector propagator keeps the same third component in the $WPV$ vertex and the $APV$ vertex and one must sum over them in the vector propagator. The vertex $AVP$ is of the type $\vec \epsilon_A\cdot \vec \epsilon_V$ \cite{Roca:2005nm} and hence one has
\begin{align}
\sum_{V \textrm{pol.}} \epsilon_i(V) \epsilon_j(V)\epsilon_j(A)= \delta_{ij} \epsilon_j(A)
=\epsilon_i(A)
\end{align}
and then the polarization of the axial-vector meson plays the same role as the polarization of the vector meson in the $WPV$ vertex and one evaluates $\overline{\sum} \sum \left|t\right|^2$ in the same way as in  Eq.~\eqref{eq:ff2}.

Taking equal masses for the mesons of the same isospin multiplet, the $G$ functions are the same for $M_1 M_2$ independent of the charges and then one can see from Tables \ref{tab:ha1} and \ref{tab:ga1} that the $M_0$ contribution cancels  for the $a_1$ ($G=-)$, ($h'_i$ coefficient), but not the $N_i$ contribution, ($\bar h'_i$ coefficient), as it should be, since we saw that $M_0$ has positive $G$-parity and $N_i$ negative $G$-parity. Conversely, in the case of the $b_1$ ($G=+$) it is the $N_i$ part that cancels, (see Tables \ref{tab:hb1} and \ref{tab:gb1}), as it should be. We see now that the change of sign in $N_i$ implied by Eq.~\eqref{eq:M0Nnu}
is essential to conserve $G$-parity in the final state interaction of the mesons. In the case of the $K_1(1270)$ resonances there is no well-defined $G$-parity but the change of sign
implied by Eq.~\eqref{eq:M0Nnu}  produces particular signs in some channels, which is important for the interference of the different contributions.

To finish the formalism, for the case of the coalescence production that we study, meaning production of the resonance independently of its decay, we have only two particles in the final state and we have the following expression for the width of the $\tau\to \nu A$ decay:

  \begin{align}\label{eq:GtauAnu}
\Gamma(\tau\to \nu A) =  \frac{2\,m_\tau 2\, m_\nu}{8\pi} \frac{1}{m^2_\tau}\, p_\nu \, \overline{\sum} \sum \left|t\right|^2 \,,
\end{align}
where we have to use for $\overline{\sum} \sum \left|t\right|^2$ the same expression as in
Eq.~\eqref{eq:ff2}  with the substitutions given in Eq.~\eqref{eq:hsubst} and $p_\nu$ the momentum of the neutrino in the tau rest frame and $p$ the momentum of the neutrino in the $A$ rest frame.

\section{Partial decay widths}

In order to ease comparison with experimental data we will also estimate the branching ratios to final $VP$ states. In order to do that, we can multiply the $\tau$ decay width into an axial-vector resonance by its branching ratio  into an specific $VP$ channel,

\begin{align}
\Gamma(\tau\to\nu_\tau A\to \nu_\tau VP)= \Gamma(\tau\to\nu_\tau A)	\frac{\Gamma(A\to VP)}{\Gamma_A} \, ,
\label{eq:tauVP}
\end{align}
with
\begin{equation}
\Gamma(A\to VP)=\frac{|g_{A,VP}|^2}{8\pi M_A^2}q.
\label{eq:GAVP}
\end{equation}
where $g_{A,VP}$ are the couplings of the axials to the specific final $VP$ channel, (see Tables~\ref{tab:ga1}-\ref{tab:gK1}).

In order to take into account the finite width of the axial and the vector mesons,
 we fold Eq.~\eqref{eq:tauVP} with  their corresponding mass distributions provided by the spectral functions of the axial,  $\rho_A(s_A)$, and  vector meson $\rho_V(s_V)$,

\begin{align}
&\Gamma(\tau\to\nu_\tau A\to \nu_\tau VP)= \nn \\
&\frac{1}{{\cal N}}
\int_{(M_A-2\Gamma_A)^2}^{(M_A+2\Gamma_A)^2} ds_A
\int_{(M_V-2\Gamma_V)^2}^{(M_V+2\Gamma_V)^2} ds_V \nn \\
&\cdot\rho_V(s_V) \rho_A(s_A) \Gamma(\tau\to\nu_\tau A\to \nu_\tau VP)(\sqrt{s_A},\sqrt{s_V})
 \nn \\
&\cdot\Theta(\sqrt{s_A}-\sqrt{s_V}-M_P)\Theta(m_\tau-\sqrt{s_A})
,
\label{eq:convo}
\end{align}

\noindent
where $\Theta$ is the step function and $\Gamma_A$
and $\Gamma_V$ are the axial and vector mesons total width.
In Eq.~(\ref{eq:convo}), ${\cal N}$ is the normalization
of the spectral distribution:
\begin{align}
{\cal N} =
\int_{(M_A-2\Gamma_A)^2}^{(M_A+2\Gamma_A)^2} ds_A
\int_{(M_V-2\Gamma_V)^2}^{(M_V+2\Gamma_V)^2} ds_V
\rho_V(s_V) \rho_A(s_A)
\end{align}
We take for the vector spectral function
\be
\rho_V(s_V)=-\frac{1}{\pi}\textrm{Im}
\left\{\frac{1}{s_V-M_V^2+iM_V\Gamma_V}\right\},
\label{eq:rho_V}
\ee
and analogously for the axial-vector one, in spite of the fact that for the axial-vector case the shape is not really a Breit-Wigner, but the approximation is good enough given the uncertainties that we will be obtaining in the results. In Eq.~(\ref{eq:convo}) $\Gamma(\tau\to\nu_\tau A\to \nu_\tau VP)$ in the integrand is to be understood as the $\Gamma(\tau\to\nu_\tau A\to \nu_\tau VP)$ explained before but substituting everywhere $M_A\to\sqrt{s_A}$ and $M_V\to\sqrt{s_V}$, (except in the $VP$ loop functions, $G$, which already had its own consideration of the finite vector meson widths  \cite{Roca:2005nm,Geng:2006yb}).
The convolution is specially relevant in the case where there is little phase space
for the decay or it is only possible thanks to the finite width of the
particles.

\section{Results}

We can take one of the branching ratios $\tau\to\nu_\tau A$  to get  the global unknown constant in Eq.~\eqref{eq:LQ}. For this we take the width $\tau\to\nu_\tau a_1(1260)$ which can be estimated using the following experimental information: Although the
 $\tau^-\to\nu_\tau \pi^+ \pi^- \pi^-$ decay is well studied experimentally, the separation of the axial-vector contribution to the rate is not done since there is interference with non resonant terms. However, if we look up in the PDG \cite{pdg,Schael:2005am} we find the information

 \begin{align}
BR(\tau^-\to \nu_\tau a_1(1260)\to \nu_\tau \pi^-\gamma)=(3.8\pm1.5)\times10^{-4}
\label{eq:BRexp}
 \end{align}
Together with the other information  used in \cite{pdg,Schael:2005am} in this analysis
\be
BR(a_1(1260)\to\gamma\pi^-)= (2.1\pm 0.8)\times 10^{-3}
\label{eq:BR2}
\ee
thus gives
 \begin{align}
BR(\tau^- \to \nu_\tau a_1(1260))=(18\pm 7)\times 10^{-2},
\label{eq:BRa1exp}
 \end{align}
where we keep the same relative error as in Eq.~\eqref{eq:BRexp}, about 40$\%$,
which, according to the analysis of  \cite{pdg,Schael:2005am}, already accounts for the error in Eq.~\eqref{eq:BR2}. The result of
Eq.~\eqref{eq:BR2} is in good agreement with the theoretical evaluation in \cite{volkovparti}
of $14\%$.
Although this formation is not necessary for the evaluation done here, concerning the nature of the $a_1(1260)$ as a dynamically generated resonance, evaluations of its radiative decay have been done assuming that nature of the $a_1(1260)$, and using the formalism of the local hidden gauge approach, and a qualitative agreement with data is obtained \cite{Roca:2006am,Nagahiro:2008cv}, which is improved if some extra genuine component for the $a1_(1260)$ is considered, as done in \cite{Nagahiro:2011jn}. Uncertainties from this source can be accomodated within the large errors that our results have.

Normalizing our results to the $a_1(1260)$ production width of Eq.~\eqref{eq:BRa1exp},
  Eq.~\eqref{eq:GtauAnu} leads to the results shown in Table~\ref{tab:resultsBR}.

%\begin{widetext}
\begin{center}
\begin{table}[h!]
\begin{center}
\caption{\small{Branching ratios (in $\%$) for creation of the different axial-vector resonances. The $\tau^-\to\nu_\tau a_1(1260)$ is fixed to the experimental result as reference.}}
\begin{tabular}{|l c|}
\hline
 Decay channel                   & BR $(\%)$        \\ \hline
 $\tau^-\to\nu_\tau a_1(1260)$   & $18\pm 7$ (exp)    \\
 $\tau^-\to\nu_\tau b_1(1235)$   & $10\pm 4$     \\
 $\tau^-\to\nu_\tau K_1(A)$      & $0.63\pm 0.25$    \\
 $\tau^-\to\nu_\tau K_1(B$       & $0.65\pm 0.26$    \\ \hline
\end{tabular}
\label{tab:resultsBR}
\end{center}
\end{table}
\end{center}
%\end{widetext}

In Tables~\ref{tab:resultsBRVP} and \ref{tab:resultsBRVPK1} we show the branching ratios (in $\%$) for each $VP$ state appearing in the coupled channels in the process $\Gamma(\tau\to\nu_\tau A\to \nu_\tau VP)$.

\begin{widetext}
\begin{center}
\begin{table}[h!]
\begin{center}
\caption{\small{Branching ratios (in $\%$) for the $\Gamma(\tau\to\nu_\tau A\to \nu_\tau VP)$ process for the $I=1$ intermediate axial-vector resonances. The results have an uncertainty of $40\%$.}}
\begin{tabular}{|c c c c c c c|}
\hline
 Decay channel                                  & $\rho^0\pi^-$ & $\rho^-\pi^0$ & $\omega\pi^-$ & $\rho^-\eta$   & $K^{*0} K^{-} $   &  $K^{*-} K^{0} $       \\ \hline
 $\tau^-\to\nu_\tau a_1(1260)\to \nu_\tau VP$   &  2.7          & 2.7          &       --      &  --            & 0.03              &    0.03                 \\
 $\tau^-\to\nu_\tau b_1(1235)\to \nu_\tau VP$   &   --          &  --           &       2.3     &   0.70          & 0.23               &    0.23                 \\\hline
\end{tabular}
\label{tab:resultsBRVP}
\end{center}
\end{table}
\end{center}
\end{widetext}

\begin{widetext}
\begin{center}
\begin{table}[h!]
\begin{center}
\caption{\small{Branching ratios (in $\%$) for the $\Gamma(\tau\to\nu_\tau A\to \nu_\tau VP)$ process for the $I=1/2$ intermediate axial-vector resonances. The results have an uncertainty of $40\%$.}}
\begin{tabular}{|c c c c c c c c|}
\hline
 Decay channel                             & $K^{*-}\pi^{0} $ & $\bar K^{*0}\pi^- $ &$\rho^0 K^- $ &$\rho^- \bar K^0 $ &$\omega K^-$ & $\phi K^-$ &$K^{*-}\eta $  \\ \hline
 $\tau^-\to\nu_\tau K_1(A)\to \nu_\tau VP$ &  0.12            & 0.23             &      0.005      &  0.010           & 0.007      &  0      & 0          \\
 $\tau^-\to\nu_\tau K_1(B)\to \nu_\tau VP$ &  0.019           & 0.037             &     0.085        & 0.17             & 0.012       &    0   &0.007          \\\hline
\end{tabular}
\label{tab:resultsBRVPK1}
\end{center}
\end{table}
\end{center}
\end{widetext}

The results of Tables~\ref{tab:resultsBRVP} and \ref{tab:resultsBRVPK1} are illustrative, indicating the channels where one expects larger partial decay widths. In this sense, as usual, the $a_1(1260)$
 production has
 to be searched for in the $\rho\pi$ mode, and the $b_1(1235)$ in the $\omega \pi$ mode. We get a branching ratio
 $
 \Gamma(\tau^-\to\nu_\tau b_1^-(1235)\to \nu_\tau \omega\pi^-)= (2.3 \pm 1.7)\%.
 $
 As we mentioned above there are no data for this decay mode but the PDG reports a branching ratio for $ \tau^-\to\nu_\tau  \omega\pi^-$ of $(1.95\pm 0.06)\%$.
 Assuming that this decay mode is dominated  by the $b_1(1235)$ one finds good agreement within errors of these results. It would be most interesting to see if the $\omega\pi^-$ mass distribution shows indeed the $b_1(1235)$ peak.

 As far as  the two $K_1(1270)$ states is concerned, as in \cite{Geng:2006yb} and other works that studied the separation of these two states using different reactions \cite{Dai:2018zki,Wang:2019mph,Wang:2020pyy}, the suggestion is always the same: to measure the $K^*\pi$ decay mode to see the $K_1(A)$ state and $\rho K$ to measure the $K_1(B)$ state. So far experiments look at the $\bar K\pi\pi$ invariant mass distribution, which contains both $K^*\pi$ and $\rho K$.

\section{Conclusions}
  We have studied the reaction $\tau \to \nu_\tau A$, with $A$ an axial-vector resonance. We find that for the Cabibbo favored decay mode, only the $a_1(1260)$ and $b_1(1235)$ resonances are produced, but the Cabibbo suppressed mode produces two $K_1(1270)$ resonances which have been predicted before. We use an approach in which the axial-vector resonances are dynamically generated from the interaction of pseudoscalar and vector mesons. For the interaction we use the chiral unitary approach. The unitarization in coupled channels of the interaction of these channels, with the only input of the lowest order $VP$ Lagrangians, produces poles for the resonances from where we extract the residues, and hence the couplings of the resonances to the different channels, which are basic ingredients in the present quantitative calculation. The weak interaction part is handled by a direct evaluation of the weak matrix elements at the quark level and using the $^3P_0$ model to hadronize the primary $q \bar q$ pair formed in the tau decay. We evaluate decay rates relative to one of the decay mode, for which we take experimental data on the $\tau \to \nu_\tau a_1(1260)$. We obtain rates for the $b_1(1235)$ production which are similar to those of the $a_1(1260)$ production and suggest to see this mode in the  $\tau^- \to \nu_\tau\omega \pi^- $ decay.
   For the case of the two $K_1(1270)$ states we show that they should be looked upon in different channels, the lower mass state should be seen in the $\bar K^* \pi$ mode, while the $K_1$ state of higher mass should appear in the $\rho K$ decay mode. We note that experiments looking for this resonance measure so far the $\bar K\pi\pi$ invariant mass that contains both the $\bar K^* \pi$ and the $\rho K$ modes.  A separation of these two channels should show two different peaks with a different width, as predicted by the theory and confirmed with other experiments \cite{Geng:2006yb}. Collecting more statistics in present facilities and with the advent of planned ones, the Hefei project in China \cite{luoxu}, another one in Novosibirsk \cite{Eidelman:2015wja}, and the Belle II update \cite{Kou:2018nap}, there will be opportunities to have a look at the issues discussed here and learn more about important aspects of Hadron dynamics, in particular the nature of hadronic resonances.

\section{Acknowledgments}
LRD acknowledges the support from the National Natural Science Foundation of China (Grant Nos. 11975009, 11575076).
This work is partly supported by the Spanish Ministerio de Economia y Competitividad and European FEDER funds under
Contracts No. FIS2017-84038-C2-1-P B and No. FIS2017-84038-C2-2-P B.
This project has received funding from the European Union's Horizon 2020 research and innovation
programme under grant agreement No. 824093 for the STRONG-2020 project.

\end{document}